\documentclass[aps,prl,notitlepage,superscriptaddress,showpacs,twocolumn]{revtex4-1}
\usepackage{graphicx,subfigure,epsfig}
\usepackage{dcolumn}
\usepackage{amssymb}
\usepackage{times}
\usepackage{amsmath}
\usepackage{amsfonts}
\usepackage{mathrsfs}
\usepackage{setspace}
\usepackage{latexsym}
\usepackage{bbm}
\usepackage{float}
\usepackage{flafter}
\usepackage{bm}
\usepackage{epstopdf}
\usepackage{hyperref}
\usepackage{color}
\usepackage{multirow}

\def\be{\begin{equation}}       \def\ee{\end{equation}}
\def\bea{\begin{eqnarray}}      \def\eea{\end{eqnarray}}
\def\ba{\begin{array}}
\def\ea{\end{array}}
\def\bnum{\begin{enumerate} }
\def\enum{\end{enumerate}}

\def\nn{\nonumber}

\def\=>{\Rightarrow}
\def\>{\rightarrow}

\def\eye2{Fathbb{I}}

\renewcommand{\>}{\rangle}

\begin{document}

\title{Exciton Condensation in Quantum Hall Bilayers at Total Filling $\nu_T=5$}

\author{Zheng Zhu}
\affiliation{Department of Physics, Massachusetts Institute of Technology, Cambridge, MA, 02139, USA}
\affiliation{Department of Physics and Astronomy, California State University, Northridge, CA, 91330, USA}
\affiliation{Department of Physics, Harvard University, Cambridge, MA, 02138, USA}
\author{Shao-Kai Jian}
\affiliation{Department of Physics, Harvard University, Cambridge, MA, 02138, USA}
\affiliation{Institute for Advanced Study, Tsinghua University, Beijing, 100084, China}
\author{D. N. Sheng}
 \affiliation{Department of Physics and Astronomy, California State University, Northridge, CA, 91330, USA}

\begin{abstract}
We study the coupled quantum Hall bilayers each at half-filled first excited Landau levels with varying  the layer distance.  Based on numerical exact diagonalization on torus, we identify 
two distinct phases separated by 
a critical layer distance $d_c$. From $d_c$ to infinite layer distance, the topological phase is smoothly connected to a direct tensor product of two Moore-Read states, while the interlayer coherence emerges at $d<d_c$ characterized by 
the $xy$ easy-plane ferromagnetic energy spectra, gapless pseudospin excitations and the finite exciton superfluid stiffness, corresponding to the exciton superfluid state. 
More interestingly, the results of  
the ground state fidelity, 
the evolution of energy spectra, and the superfluid stiffness 
indicate a possible continuous transition. Theoretically 
it can be interpreted as a topological phase transition which simultaneously changes the topology of ground state and breaks symmetry, providing an interesting  example of transitions beyond Landau paradigm.
\end{abstract}
\maketitle

\emph{Introduction.}---The multicomponent fractional quantum Hall (FQH) systems, especially quantum Hall bilayers, have triggered substantial interest  in pursuing exotic emerging quantum states~\cite{Girvin,Halperin1983}. The quantum Hall bilayers, which can be realized in single wide quantum wells or  double quantum wells~\cite{bilayer1,bilayer2}, host rich phenomena such as the Bose-Einstein condensation of the excitons~\cite{Eisenstein2004,Eisenstein2014} and   anyonic  statistics ~\cite{Laughlin83,MR1991,Greiter1991,RR1991}. The possible emerging non-Abelian physics and quantum phase transitions among the various phases 
remain not well understood
~\cite{Wen2000,Barkeshli2010,Barkeshli2011}.

In particular,  the bilayers with half filled lowest Landau level (LL) for each layer have  attracted  great interest from  both experimental measurements~\cite{Kellogg2004,Wiersma2004,Murphy1994,Giudici2010,Eisenstein2014,Narozhny2016} and theoretical  investigations~\cite{Cote1992,Wen1992, Moon1995,Bonesteel1996,Kim2001,Joglekar2001,Stern2002,Veillette2002,Simon2003,Wang2003, Doretto2006, Alicea2009,Cipri2014,Isobe2016,Sodemann2016,Potter2016,Lian2017,You2017,Milovanovic2017,
Schliemann2001,Shibata2006,Sheng2003,Park2004,Moller2008,Milovanovic2015,Zhu2017}.
The  exciton superfluid phase (or Haplerin ``111" state~\cite{Halperin1983,Yoshioka1989}) was first established experimentally at a layer distance comparable to the magnetic length~\cite{Eisenstein2014} based on a  zero-bias interlayer tunneling conductance~\cite{Spielman2000} and a vanishing Hall counterflow resistance~\cite{Kellogg2004,Tutuc2004}. Other  phases like the composite Fermi liquid (CFL) at larger distance~\cite{Halperin1993} and the novel intermediate phase~\cite{Moller2008,Milovanovic2015,Zhu2017} have also been extensively investigated. In addition, the nature of quantum phase transitions among various phases is still controversial. Inspired by the rich physics of these $\nu_T=1$ bilayers with half filled lowest LL in each layer, a natural question arises about the quantum phase diagram for the electronic systems with fully filled lowest LL and half-filled first excited LL, corresponding to the bilayers with total filling $\nu_T=5$. Each decoupled layer with filling $\nu=5/2$ is believed to be the Moore-Read (MR) state with intralayer paired composite fermions  and non-abelian Ising anyons~\cite{MR1991,Read2000,Jain2000}. When the layer distance goes to zero, the interlayer coherent state is theoretically expected though there is no experimental study presented  along this line. By tuning the layer distance, the nature of possible intermediate phase and the quantum phase transition remain unclear, which motivate our present work.

Previous theoretical studies have not reached a consistent conclusion on this problem. On one hand, the calculations based on Hartree-Fock approximation claim a transition from ``111" state  to a charge ordered state~\cite{Wang2003,Brey2000}, while the varational  and exact diagonalization (ED) calculations on sphere geometry found a bilayer phase coherent state at small layer distance and  two uncoupled 5/2 state at large layer separations by Shi \emph{et al}~\cite{Shi2008}.
Nevertheless, unbiased exact simulations for quantum states at intermediate layer distances and the quantum phase transition for torus geometry are still absent. 
Different from the sphere geometry,  there is no orbital number shift on torus and the competing states with the same filling factor can be compared on an equal footing~\cite{Haldane1985}, which is more suitable for studying 
quantum phase transitions. 

In the present work, we use ED to calculate systems with up to 18 electrons on torus. 
Based on the energy spectra, pseudospin gap, exciton superfluid stiffness, the Berry curvature as well as drag Hall conductance, we identify a direct phase transition at $d_c$ between the exciton superfluid phase with interlayer coherence and the phase with strong intralayer correlations, the latter can be smoothly connected to the decoupled two copies of MR state. Here the finding of $d_c$ is consistent with previous varational calculation~\cite{Shi2008}.
Moreover, the calculation of fidelity, the exciton superfluid stiffness, the evolution of energy spectrum,
and the ground state energy derivatives indicate the transition
is continuous, which is beyond the  Landau paradigm~\cite{Wen2000, Barkeshli2010, Barkeshli2011, Landau1958, Senthil2004}. Based on the analysis of symmetries and topological orders, we propose the theoretical interpretations of such a transition as exciton condensation which simultaneously breaks $U(1)\times U(1)$ symmetry and changes the topology. The exciton condensation leads to the $C=2$ topologically ordered state in Kitaev's notation~\cite{Kitaev2006}, which is consistent with the ``111" state.

\emph{Model and Method.}--- We consider the $\nu_T=5/2+5/2$ bilayer electronic systems subject to a perpendicular magnetic field. We neglect the width of these two identical layers  and put them on torus spanned by vectors $\mathbf{L_x}$ and $\mathbf{L_y}$. The orbital number (or flux number) in each layer $N_\phi$ is determined by the area of torus, i.e., 
$|\mathbf{L_x}\times \mathbf{L_y}|=2\pi N_\phi$. In the absence of the interlayer tunneling, this system with fully polarized spins can be described by the projected Coulomb interaction, which reads
 \begin{equation}\label {Ham}
V =\frac{1}{N_\phi} \sum\limits_{i < j,\alpha ,\beta } {\sum\limits_{{\bf{q}},{\bf{q}} \ne 0} {{V_{\alpha \beta }}\left( q \right)} } {e^{ -\frac{q^2}{2}}}L^2_n[\frac{q^2}{2}]{e^{i{\bf{q}} \cdot \left( {{{\bf{R}}_{\alpha ,i}} - {{\bf{R}}_{\beta ,j}}} \right)}}.
  \end{equation}
Here, $\alpha (\beta)=1,2$ denote two layers or, equivalently, two components of a pseudospin-$1/2$. $q$ =$|{\bf{q}}|$ =$ \sqrt {q_x^2 + q_y^2} $, $V_{11}(q)= V_{22}(q)={e^2}/({\varepsilon q})$ and $V_{12}(q)=V_{21}(q)= {e^2} /({\varepsilon q})\cdot e^{-qd}$  are the Fourier transformations of the intralayer and interlayer Coulomb interactions, respectively. $d$ represents the distance between two layers in the unit of magnetic length $l_B$. $L_n(x)$ is the Laguerre polynomial with Landau level index $n$ and $\bf{R}_{\alpha,i}$ is the  guiding center coordinate of the $i$-th electron in layer $\alpha$. In particular, we consider rectangular unit cells with $L_x=L_y=L$  when using ED to study the energy spectrum and state information on torus.

\begin{figure}[tbp]
\begin{center}
\includegraphics[width=0.45\textwidth]{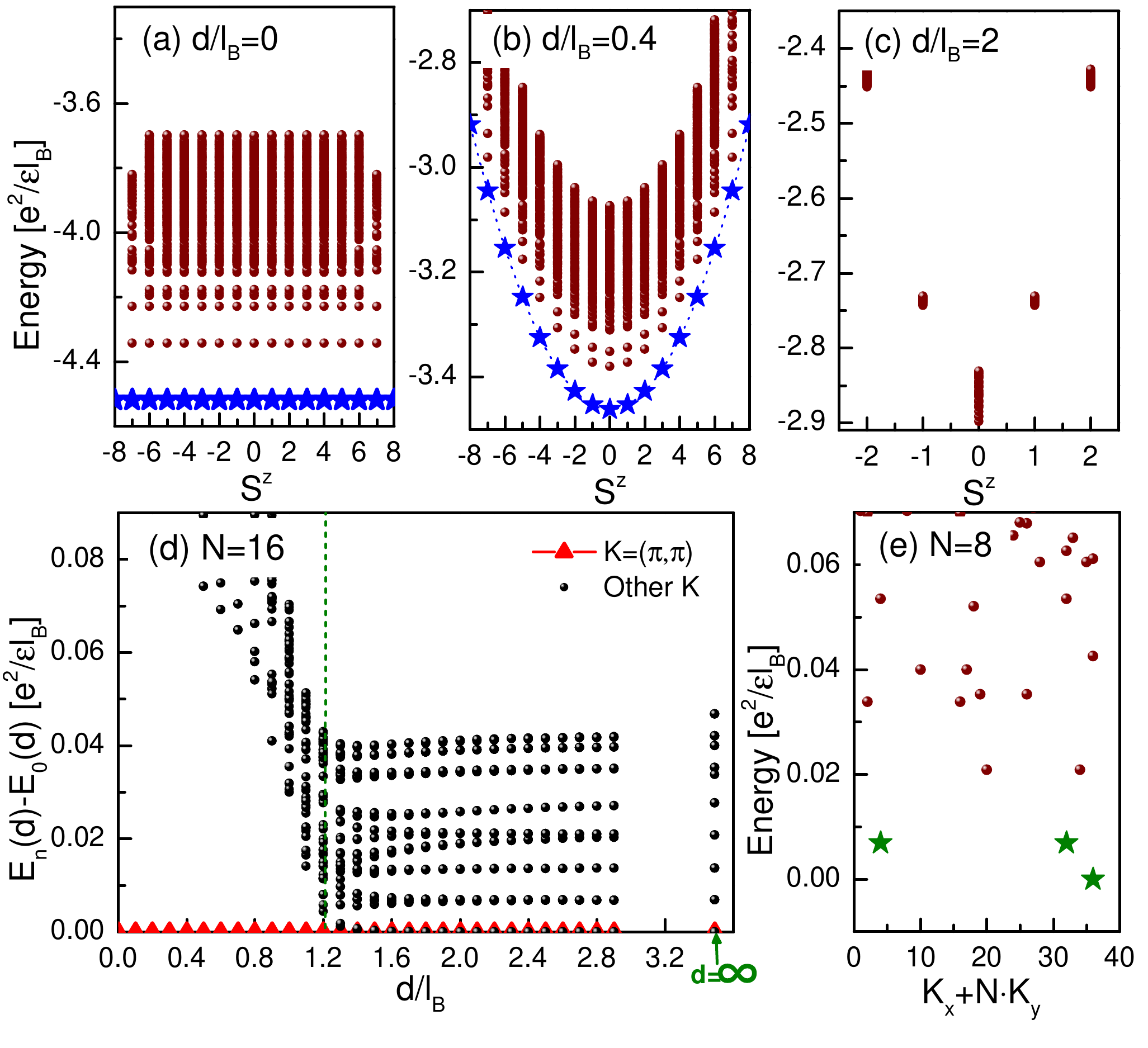}
\end{center}
\par
\renewcommand{\figurename}{Fig.}
\caption{(Color online) The energy spectra of different pseudospin $S_z$ sectors at layer distance (a) $d/l_B=0$, (b) $d/l_B=0.4$, (c) $d/l_B=2$. (d) The low-lying energy spectra as a function of layer distance $d/l_B$. Here,  the total electron number$N=16$ and each layer has equal number of electrons.  (e) The energy spectrum of single layer $N=8$ system at n=1LL, the green stars highlight the topological sectors of MR state.   }
\label{Fig:EK}
\end{figure}

 \emph{Energy Spectra.}---Without  the interlayer tunneling, the bilayer system has separate conservations for the electron number in each layer, which allows us to label eigenstates by pseudospin $S_z$ defined as $S_z\equiv (N^\uparrow-N^\downarrow)/2$, where $N^\uparrow$ and $N^\downarrow$ denote the number of electrons for the  top and bottom layers, respectively. Then we can study the energy spectra  by targeting different pseudospin sectors.  Here, the energy shift $d\cdot S^2_z/N_\phi$ induced by the imbalance of charge in two layers~\cite{MacDonald1990} has been considered. When the layer distance goes to zero, as shown in Fig.~\ref{Fig:EK} (a), the lowest energies in each pseudospin $S_z$ sector are exactly degenerate, indicating that we have not only conserved $S_z$ but also full 
$SU(2)$ symmetry.
This spectrum is consistent with the exciton condensed ``111" state,  with spontaneous ferromagnetization which can be seen from the ground state spin degeneracy.
 However, when the layer distance is finite but small enough, as shown in the Fig.~\ref{Fig:EK} (b), our data shows the nondegenerate
ground state located in the  $S_z=0$ sector, and the low energy excitations are pseudospin excitations among different $S_z$ sectors, which can be fitted into $\Delta E=E(S_z)-E(S_z=0)=\alpha S_z^2$. These facts indicate that the ground state is an $xy$ easy-plane ferromagnet instead of Ising ferromagnet,  and the interlayer correlations dominate the low energy physics for small $d$. 
Physically, an electron in one layer is bound to a hole in the other layer forming an exciton at $d=0$, then  the bilayer system can be mapped into a monolayer at $\nu=1$ for the first excited Landau level.  When $d$ is finite but smaller than a critical value, a difference between the interlayer and intralayer Coulomb interaction breaks the pseudospin invariance  down to $U(1)$, leading to the $xy$ easy-plane pseudospin ferromagnet as indicated in Fig.~\ref{Fig:EK} (a) and (b). However, for larger  layer distance $d=2.0l_B$, the lowest energy excitations exist within the same pseudospin  $S_z$ sector [see Fig.~\ref{Fig:EK} (c)], indicating the low-lying excitations are dominated by the intralayer correlations.  These results indicate there are two distinct phases as the layer distance $d$ is varied.

The flow of low-lying energies with $d/l_B$ indeed indicates a direct transition at $d_c/l_B\approx1.2$ from exciton superfluid phase ($d<d_c$) to a phase with distinct structure of spectra ($d>d_c$) which can be smoothly connected to the decoupled two copies of MR state at $d/l_B=+\infty$  [see Fig.~\ref{Fig:EK}(a)].
Figure \ref{Fig:EK} (e) shows the energy spectrum of each decoupled layer with 8 electrons,  where the threefold degeneracy (in addition to the twofold center of mass degeneracy) in momentum sectors $(K_x,K_y)/(2\pi/N)=(N/2,N/2),(0,N/2),(N/2,0)$ occurs 
supporting that each decoupled layer is indeed in the MR state. When coupling two layers together, we identified a 36-fold near degeneracy of two copies of MR state at $d>d_c$ side.
 
\begin{figure}[tbp]
\begin{center}
\includegraphics[width=0.45\textwidth]{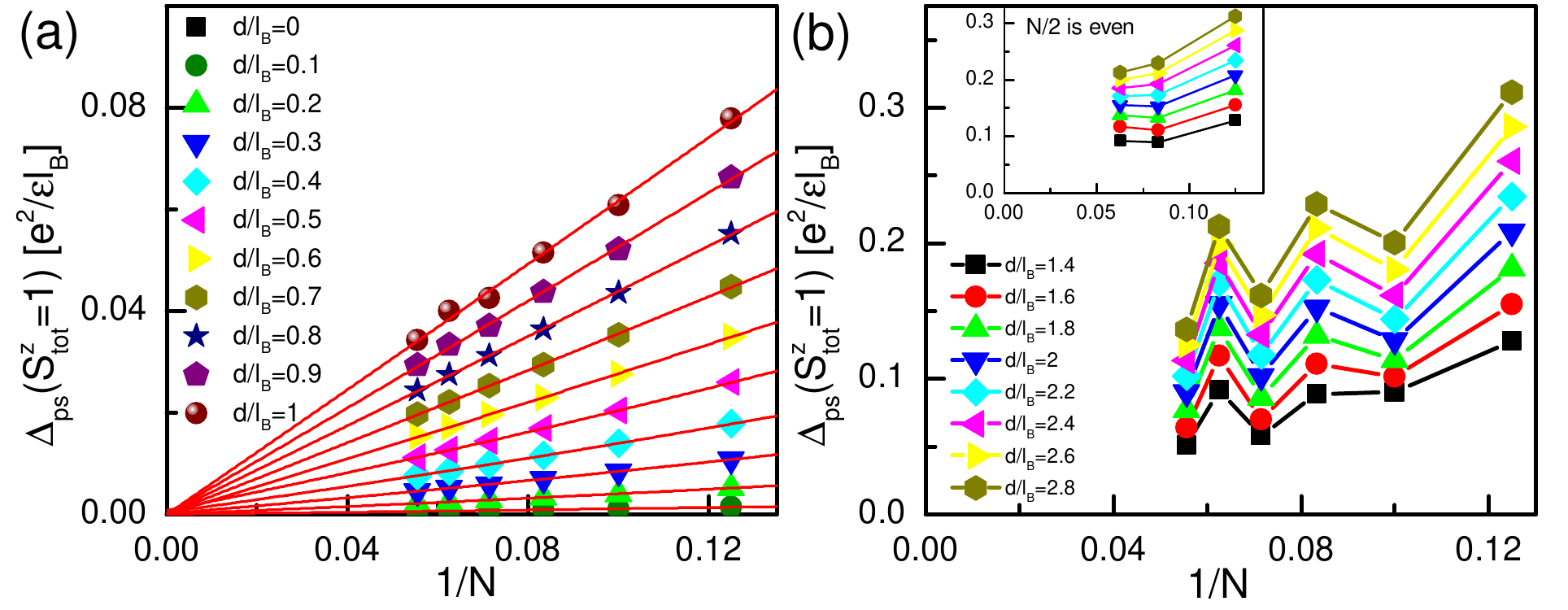}
\end{center}
\par
\renewcommand{\figurename}{Fig.}
\caption{(Color online)  The single pseudospin excitation gap $\Delta_{ps}$ for the phase at $d<d_c$ (a) and $d>d_c$(b). The finite size scaling of $\Delta_{ps}$  using parabolic function indicates the gapless nature at $d<d_c$ (a), while $\Delta_{ps}$ displays even-odd effect at $d>d_c$ (b), the inset of (b) show the spin gap for the systems with even number of particles in each layer.}
\label{Fig:SpinGap}
\end{figure}

 \emph{Pseudospin Excitations.}---From the energy spectra we identify a single phase transition at $d_c/l_B\approx1.2$ without 
energy  level crossing. 
Below we characterize the transition  from the perspective of low energy excitations.
We directly calculate the pseudospin excitation gap, which physically measures the energy cost when flipping the pseudospin
of one particle. The pseudospin gap is defined as $ \Delta_{ps}(d)\equiv E_0(N_\uparrow, N_\downarrow,d)-E_0(N/2,N/2,d)+d\cdot S^2_z/N_\phi$, where $N_\uparrow=N/2+S_z $ and $N_\downarrow=N/2-S_z$. 
As shown in Fig.~\ref{Fig:SpinGap} (a), the finite size scaling of $\Delta_{ps}(d)$ for  $S_z=1$  shows the excitation gap goes  to  zero in the  thermodynamic limit for $d/l_B \lesssim 1.2$, which is consistent with interlayer coherence of pseudospins or the existence of the Goldstone mode in exciton superfluid phase. However,  for $d/l_B\gtrsim 1.2$, the $S_z=1$ pseudospin excitation displays significant even-odd effect determined by the electron number in each layer, as shown in Fig.~\ref{Fig:SpinGap} (b).   For the systems with even number of electrons in each layer,  flipping a single pseudospin costs finite energy [see the inset of Fig.~\ref{Fig:SpinGap} (b)]  while the energy cost vanishes when the electron number in each layer is odd. This even-odd effect suggests the existence of intralayer paired composite fermions. 
Therefore, with the decrease of the layer distance, the transition at $d_c$ is characterized by the closing of pseudospin gap and the disappearance of even-odd effect. 
Furthermore, we will show below that the interlayer coherence  immediately establishes in the gapless phase at $d\leqslant d_c$, leading to the exciton superfluidity.

\begin{figure}[tbp]
\begin{center}
\includegraphics[width=0.45\textwidth]{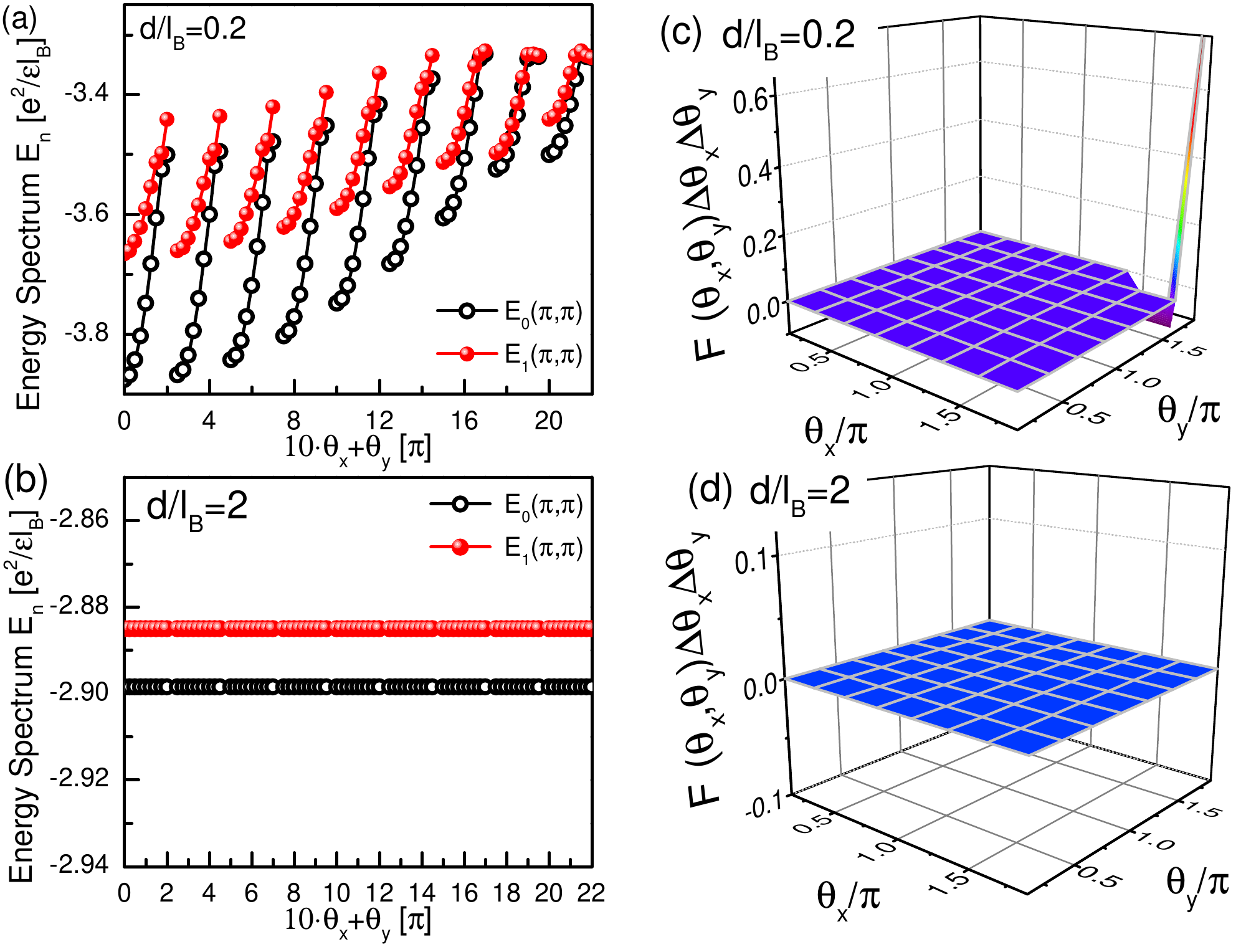}
\end{center}
\par
\renewcommand{\figurename}{Fig.}
\caption{(Color online)  The energy flow of  $N=16$ system with twisted  boundary phases for  $d/l_B$=0.2 (a),  $d/l_B$=2 (b). Here, we map the phase $\theta_x, \theta_y$ into one-dimensional quantity $\theta\equiv10\theta_x+\theta_y$ for convenience of plotting.  (c) and (d) are the Berry curvature $ F(\theta _x^\alpha,\theta _y^\beta)$ for $d/l_B$=0.4 (c) , $d/l_B$=2 (d). Here, $\Delta \theta_x \Delta \theta_y$ is the area of the mesh in phase space. }
\label{Fig:E_Twist}
\end{figure}

\begin{figure}[t]
\begin{center}
\includegraphics[width=0.45\textwidth]{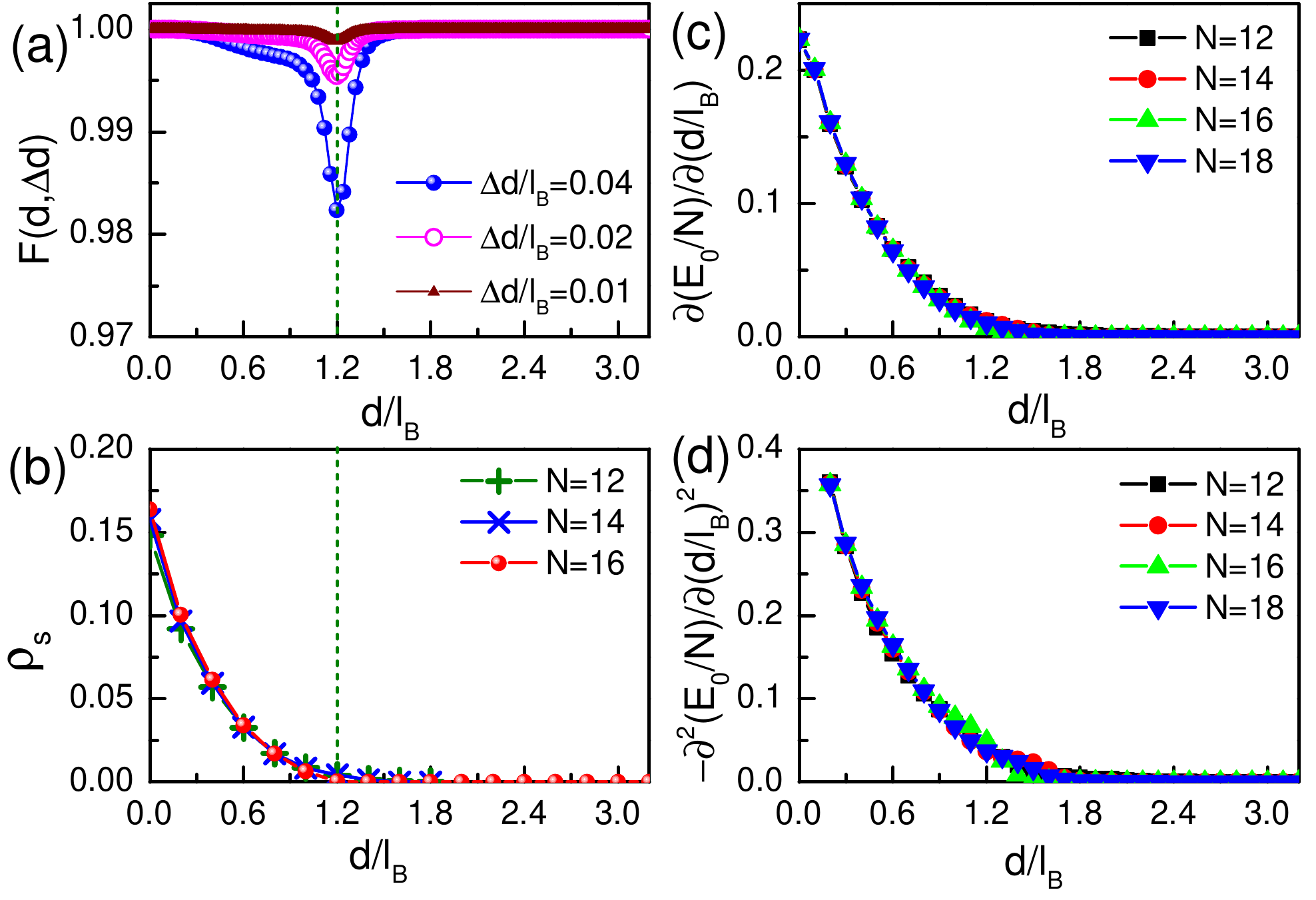}
\end{center}
\par
\renewcommand{\figurename}{Fig.}
\caption{(Color online) (a) The Fidelity of $N=16$ systems with layer distances and different interval of parameters $\Delta d$. (b) The exciton superfluid stiffness $\rho_s$  as a function of layer distance. The first-order (a) and second-order (b) derivative curves of  ground-state energy $E_0/N$ as a function of  layer distance $d/l_B$.  }
\label{Fig:Transition}
\end{figure}

 \emph{Exciton Superfluid Stiffness.}--- To study the interlayer coherence, we add twisted  boundary  phases $0\leq\theta _\lambda ^\alpha\leq2\pi $ along $\lambda$ direction ($\lambda=x$ or $y$) in the layer $\alpha$, and study the energy evolution  in order to get the exciton superfluid stiffness $\rho_s$. Physically  
imposing opposite boundary phases for two layers plays a similar
 role as the counterflow experiments, where the longitudinal counterflow conductivity indicates the superfluidity. 
Fig.~\ref{Fig:E_Twist} (a) and (b) show the energy flow of the lowest two states in the same momentum sector $(K_x,K_y)=(\pi,\pi)$ with twisted phases.
 The exciton superfluid stiffness $\rho_s$, can be obtained by fitting the energy flows according to~\cite{Sheng2003}
  \begin{equation}\label{Eq:rho_s}
 E(\theta_t)/{|{\bf L_x}\times{\bf L_y}|}=E(\theta_t=0)/A+\frac{1}{2}\rho_s \theta^2_t+O(\theta^4_t),
 \end{equation}
 where $E(\theta_t)$ is the ground-state energy with twisted (opposite) boundary phases $\theta_t$ between two layers. Based on  Eq.~\ref{Eq:rho_s} we fit the energy curve using the quadratic function to extract  $\rho_s$.  As shown in Fig.~\ref{Fig:E_Twist} (a), we have finite exciton superfluid stiffness when $d<d_c$ , while  $\rho_s=0$ at $d>d_c$ side due to the totally flat energy curve against twisted phases, indicating the vanish of superfluidity. The quantitative evolution of the superfluid stiffness $\rho_s>0$ with the layer distance will be discussed later in Fig.~\ref{Fig:Transition} (b)  to address the precise nature of the quantum phase transition.

 \emph{Berry curvature and drag Hall conductance.}---We have identified that $d_c$ separates two phases with and without interlayer coherence. Physically, the interlayer correlations can also be detected by the drag Hall conductance, which can be calculated  by integrating the Berry curvature $ F(\theta _x^\alpha,\theta _y^\beta)$. Here we study the Berry curvature based on the standard techniques~\cite{Thouless1982,Niu1985,Wen1991,Yang2001,Sheng2003,Sheng2006,Sheng2011} using twisted boundary phases. Figures~\ref{Fig:E_Twist} (c) and (d) show the Berry curvatures at the $d<d_c$ and $d>d_c$ , respectively, by applying $\theta^1_x=\theta_x$, $\theta^2_x=0$ and $\theta^1_y=0$, $\theta^2_y=\theta_y$ (obtaining the off-diagonal term of the Chern number matrix) for the lowest energy state in the sector $(\pi,\pi)$. For phase $d<d_{c }$ [see Fig.~\ref{Fig:E_Twist} (c)], the Berry curvature exhibits strong fluctuations, with a singular large value near the twisted phase point $(2\pi, 2\pi)$, which is caused
by the gap closing between the ground state and the first excited state near this point.
At $d>d_c$ side, the Berry curvature is near flat without any singularity [see Fig.~\ref{Fig:E_Twist} (d)]. Physically, a gapped state has a well-defined Berry curvature and thus well-defined Chern number, while a gapless state has singularities in Berry curvature due
to the energy level crossing.

The integral  over the boundary phase unit cell leads to the topological Chern number~\cite{Thouless1982,Niu1985,Wen1991,Yang2001,Sheng2003,Sheng2006,Sheng2011} matrix $C_{\alpha,\beta}=1/2\pi\int d\theta _x^\alpha d\theta _y^\beta F(\theta _x^\alpha,\theta _y^\beta)$.  Numerically,  the Hall conductances in the layer symmetric channel $C^c (e^2/h)$ and antisymmetric channel $C^s (e^2/h)$ can be calculated by applying common and opposite boundary phases on two layers, respectively. The drag Hall conductance $\sigma^d_{xy}=(C^c-C^s)(e^2/{2h})$ can be obtained directly by  twisting boundary phases along $x$ direction in one layer and along $y$ direction in another layer~\cite{Sheng2003}. For the exciton superfluid phase at $d<d_c$, the absence of well defined drag Hall conductance is consistent with the absence of well defined $C^s (e^2/h)$ due to gapless pseudospin excitations, though the $C^c (e^2/h)$ is quantized ensured by the charge gap. For the phase at $d>d_c$, both charge and pseudospin excitation are gapped,
 and the Berry curvature is near flat without any singularity and its integral leads to zero drag Hall conductance, indicating the Hall conductances are equal in both layer symmetric and antisymmetric channels, i.e., $C^c =C^s$.

\emph{Continuous phase transition.}---Since the exciton superfluid phase and two copies of MR phase  have different symmetries and topological orders, a direct continuous transition is beyond the  Landau paradigm.
From the energy spectra in Fig.~\ref{Fig:EK} (d), the level crossing is absent in the vicinity of the critical distance $d_c$, indicating a continuous transition.
We further probe the nature of such a transition by calculating the ground state fidelity, superfluid stiffness as well as the ground-state energy derivatives. The fidelity is defined by the wave function overlap between the ground state at $d-\Delta d$ and $d$, i,e, $F(d,\Delta d) = \left| {\left\langle \Psi(d-\Delta d)\right|\left. \Psi(d) \right\rangle } \right|$, which has been shown to be a good indicator
to distinguish continuous transition from first-order transition for both symmetry-breaking and topological phase transitions~\cite{Zanardi2006,Gu2010}. As shown in Fig.~\ref{Fig:Transition} (a),  we find the ground-state fidelity displays a single weak dip at the critical distance $d_c$ instead of showing a sudden jump.
In addition, as shown in Fig.~\ref{Fig:Transition} (b), the exciton superfluid stiffness $\rho_s$ is finite at $d<d_c$, but smoothly decreases with the increase of the layer distance, and becomes vanishingly small after the transition. Figures~\ref{Fig:Transition} (c) and (d) show the first-order and second order derivatives of the ground-state energy, which are both  smooth functions of  layer distances. Thus the numerical evidence indicates the direct transition between these two phases might be continuous, which is beyond the Landau paradigm.

\emph{Field theory of transition and exciton condensation.}---Here, we provide a possible scenario of the observed transition. We consider the electron to be fractionalized into a boson and a fermion with emergent $u(1)_i$ gauge field at each layer, i.e., $c_i = b_i \psi_i $, where $i=\uparrow,\downarrow$ denotes two layers. While the $\psi_i$ only carries $u(1)_i$ charge, $b_i$ carries both emergent $u(1)_i$ and global $U(1)_i$ charge (corresponding to the charge conservation at each layer).
To obtain MR state at each layer, pairs of fermions form $p+ip$ superconductor~\cite{MR1991, Read2000},
while pairs of bosons form $\nu=1/8$ state called $u(1)_8$ state~\cite{Bonderson2013, Chen2014, Seiberg2016}. The effective theory is
\bea
	\mathcal{L} &=& \sum_{i=\uparrow, \downarrow}  \Big( \frac{8}{4\pi} \alpha_i d\alpha_i  + \frac{2}{2\pi} ( e A_i + a_i) d\alpha_i - \frac{2 e^2}{4\pi} A_i d A_i \nn \\
	&& + \Psi_i^\dag [i\partial_0- a_{i,0}+ h_i(\vec p + \vec a_i)]  \Psi_i \Big),
\eea
where $a_{i,\mu}$ is the emergent gauge field from fractionalization, and $\alpha_{i,\mu}$ characterizes the $u(1)_{8}$ state at $i$-th layer. $ada$ is a short hand notation of the Chern-Simons term $\epsilon^{\mu\nu\rho} a_\mu \partial_\nu a_\rho$~\cite{Wen1995, Wen2004}. The first two terms correspond to $u(1)_8$ state, and the third term characterizes the Hall response of the filled lowest LL. Integrating out $\alpha_i$ field
gives rise to the quantized Hall conductivity $\sigma_{xy}=\frac52 \frac{e^2}{h}$ for each layer. In the second line, $\Psi_i(p) = (\psi_i(p), \psi^\dag_i(-p))^T$ is the Nambu spinor, $ h_i(\vec p) =\Big(\frac{p_x^2 + p_y^2}{2m} -\mu \Big) \sigma^z + \Delta_i (p_x \sigma^x + p_y \sigma^y) $ is the Bogoliubov-de-Gennes (BdG)
Hamiltonian of $p+ip$ superconductors (SC) at $i$-th layer with Pauli matrix $\sigma$ acting on Nambu space, and $\Delta_i$ denotes the pairing condensate. $m>0$ and $\mu$ are the effective mass and chemical potential of fractionalized fermion. When $\mu>0$, the $p+i p$ SC is in the topological phase with BdG Chern number $C=1$ at each layer~\cite{Read2000, Kitaev2006, Alicea2012}.

The transition to ``111" state is described by interlayer exciton condensation $ \langle c_\uparrow c_\downarrow^\dag \rangle=\langle b_\uparrow b_\downarrow^\dag \rangle \langle \psi_\uparrow \psi_\downarrow^\dag \rangle \ne0$, which simultaneously breaks $S_z$ conservation and leads to $C=2$ topological order~\cite{Kitaev2006}. It is possible that $\langle \psi_\uparrow \psi_\downarrow^\dag \rangle$ becomes nonzero breaking the residue $Z_2^\uparrow \times Z_2^\downarrow$ of the emergent $u(1)_\uparrow \times u(1)_\downarrow$ symmetry before exciton condensation, but exciton inducing the condensation of $\langle \psi_\uparrow \psi_\downarrow^\dag \rangle$ is not a fine tuned result, and indeed the numerical results show a single transition. In the presence of interlayer coherence, the Hamiltonian of fractionalized fermion is
$
	H = \left( \ba{cccc} h_\uparrow & h_{\uparrow\downarrow}
			\\ h_{\uparrow\downarrow}^\dag & h_{\downarrow} \ea \right),
$
where $h_{\uparrow\downarrow} = \text{diag} (\Phi, -\Phi^\ast)$ with $\Phi=\langle \psi_\uparrow \psi_\downarrow^\dag \rangle$.
The BdG Chern number
is the sum of two layers $C=1+1=2$. The  topological order of the $C=2$ state is Abelian, which can be captured by a Chern-Simons term~\cite{Sodemann2016, Kitaev2006}:
\bea
	\mathcal{L} &=& \sum_i \Big[\frac8{4\pi} \alpha_i d \alpha_i + \frac2{2\pi}(e A_i + a_i) d\alpha_i   \nn\\
	&& + \frac1{2\pi} \beta_i da_i \Big]- \frac{4}{4\pi} \beta_- d \beta_-,
\eea
where $\beta_-= \frac{\beta_\uparrow-\beta_\downarrow}2$, $\beta_i$ is the dual theory~\cite{Seiberg2016, Peskin1978, Halperin1981} of Higgs field $\Delta_i$ that breaks $u(1)_i$ to $Z_2^i$, and gapped fermion part is neglected. The last term is the forecasted Chern-Simons term to capture the quartonic statistics of the  $C=2$ topological order.
Now we can integrate out $a_i$ since they are linear in the Lagrangian to get
$
	\mathcal{L} =  \sum_{ij} \beta_i K_{ij} d \beta_j +\sum_i \frac{e}{2\pi} A_i d\beta_i,
$
where $K= \left( \ba{cccc} 1 & 1 \\ 1 & 1 \ea \right)$ corresponding to the ``111" state \cite{Wen1992}. The $K$ matrix indicates a gapless gauge field, $\beta_-$, dual to Goldstone bosons originated from $S^z$ symmetry breaking.

{\it Concluding remarks.}---We have shown a direct continuous transition between a non-abelian state, which can be smoothly connected to two uncoupled MR states, and a symmetry-breaking exciton superfluid state, which exhibits gapless neutral excitations, 
in the bilayer quantum Hall system with $\nu_T=5/2+5/2 =5$. We further propose an exotic scenario of such transition, where the topology changing and symmetry breaking take place simultaneously. Our results suggest  possible  experimental detections of the exotic topological quantum transition in bilayer quantum Hall systems.

\begin{acknowledgments}
We acknowledge helpful discussions with L. Fu, I. Sodemann, B. Lian, Y. H. Zhang, Y. Z. You, and Z.-X. Liu.  Z. Z. is grateful to Liang Fu for previous collaborations on $\nu_T=1$ quantum Hall bilayers. S.-K. J. acknowledges the Tsinghua Visiting Doctoral Students Foundation. This research  is supported by the U.S. Department of Energy, Office of Basic Energy Sciences under grants No. DE-FG02-06ER46305.
\end{acknowledgments}


\begin{thebibliography}{99}

\bibitem{Girvin}S. M. Girvin and A.H. MacDonald, {\it Perspectives in Quantum Hall Effects}, edited by A. Pinczuk and S. Das Sarma (Wiley, New York, 1997).
\bibitem{Halperin1983}B. I. Halperin, Helv. Phys. Acta \textbf{56},75 (1983).


  \bibitem{bilayer1} Y. W. Suen, L. W. Engel, M. B. Santos, M. Shayegan, and D. C. Tsui, Phys. Rev. Lett. \textbf{68}, 1379 (1992);
 \bibitem{bilayer2}J. P. Eisenstein, G. S. Boebinger, L. N. Pfeiffer, K. W. West, and S. He, Phys. Rev. Lett. \textbf{68}, 1383 (1992).

 \bibitem{Eisenstein2004} J. P. Eisenstein and A. H. Macdonald, Nature \textbf{432}, 691 (2004).
 \bibitem{Eisenstein2014} J. P. Eisenstein, Annu. Rev. Condens. Matter Phys. \textbf{5}, 159 (2014), and the references therein.
 \bibitem{Laughlin83} R. B. Laughlin, Phys. Rev. Lett. \textbf{50}, 1395 (1983).
\bibitem{MR1991}  G. Moore and N. Read, Nucl. Phys. B \textbf{360}, 362 (1991).
\bibitem{Greiter1991} M. Greiter, X.-G. Wen, and F. Wilczek, Phys. Rev. Lett. \textbf{66}, 3205 (1991)
\bibitem{RR1991}N. Read and E. Rezayi, Phys. Rev. B \textbf{59}, 8084 (1999)

\bibitem{Wen2000}X.-G. Wen, Phys. Rev. Lett. \textbf{84}, 3950 (2000).
\bibitem{Barkeshli2010} M. Barkeshli and X.-G. Wen, Phys. Rev. Lett. 105, 216804 (2010).
\bibitem{Barkeshli2011}M. Barkeshli and X.-G. Wen, Phys. Rev. B 84, 115121 (2011).

 \bibitem{Wiersma2004}R. D. Wiersma, J. G. S. Lok, S. Kraus, W. Dietsche, K. von Klitzing, D. Schuh, M. Bichler, H.-P. Tranitz, and W. Wegscheider, Phys. Rev. Lett. 93, 266805 (2004).
 \bibitem{Murphy1994} S. Q. Murphy, J. P. Eisenstein, G. S. Boebinger, L. N. Pfeiffer, and K. W. West, Phys. Rev. Lett. \textbf{72}, 728 (1994).
 \bibitem{Giudici2010}P. Giudici, K. Muraki, N. Kumada, and T. Fujisawa, Phys. Rev. Lett. \textbf{104}, 056802 (2010); P. Giudici, K. Muraki, N. Kumada, Y. Hirayama, and T. Fujisawa, Phys. Rev. Lett. \textbf{100}, 106803 (2008).
 \bibitem{Kellogg2004} M. Kellogg, J. P. Eisenstein, L. N. Pfeiffer, and K.W.West, Phys. Rev. Lett. \textbf{93}, 036801 (2004); Phys. Rev. Lett. \textbf{90}, 246801 (2003).
 \bibitem{Narozhny2016} B. N. Narozhny and A. Levchenko, Rev. Mod. Phys. \textbf{88}, 025003 (2016), and the references therein.
 \bibitem{Schliemann2001} J. Schliemann, S. M. Girvin, and A. H. MacDonald, Phys. Rev. Lett. \textbf{86}, 1849 (2001). John Schliemann,Phys. Rev. B \textbf{67}, 035328(2003).
  \bibitem{Shibata2006} N. Shibata and D. Yoshioka, J. Phys. Soc. Jpn. \textbf{75}, 043712 (2006).
\bibitem{Sheng2003}D. N. Sheng, L. Balents, and Z. Wang, Phys. Rev. Lett. \textbf{91}, 116802 (2003).

\bibitem{Park2004} K. Park, Phys. Rev. B \textbf{69}, 045319 (2004).
 \bibitem{Moller2008}G. M\"{o}ller, S. H. Simon, and E. H. Rezayi, Phys. Rev. Lett. \textbf{101}, 176803  (2008),G. M\"{o}ller, S. H. Simon, and E. H. Rezayi, Phys. Rev. B \textbf{79}, 125106 (2009).
\bibitem{Milovanovic2015} M. V. Milovanovi\'c, E. Dobardzic, and Z. Papi\'{c}, Phys. Rev. B \textbf{92}, 195311  (2015).
\bibitem{Zhu2017} Z. Zhu, L. Fu, and D. N. Sheng,   Phys. Rev. Lett. \textbf{119}, 177601(2017)

\bibitem{Cote1992}R. Cote, L. Brey, and A. H. MacDonald, Phys. Rev. B \textbf{46},10239 (1992).
\bibitem{Wen1992}X.-G. Wen and A. Zee, Phys. Rev. Lett. \textbf{69}, 1811 (1992).
%
\bibitem{Moon1995}K. Moon, H. Mori, K. Yang, S. M. Girvin, A. H. MacDonald, L. Zheng, D. Yoshioka, and S.-C. Zhang, Phys. Rev. B \textbf{51}, 5138 (1995).
\bibitem{Bonesteel1996}N. E. Bonesteel, I. A. McDonald, and C. Nayak, Phys. Rev. Lett. \textbf{77}, 3009 (1996).
\bibitem{Kim2001}Y. B. Kim, C. Nayak, E. Demler, N. Read, and S. Das Sarma, Phys. Rev. B \textbf{63}, 205315 (2001).
\bibitem{Joglekar2001}Y. N. Joglekar and A. H. MacDonald, Phys. Rev. B \textbf{64}, 155315 (2001).
\bibitem{Stern2002}A. Stern and B. I. Halperin, Phys. Rev. Lett. \textbf{88}, 106801 (2002).
\bibitem{Veillette2002}M. Y. Veillette, L. Balents, and M.P.A. Fisher, Phys. Rev. B \textbf{66}, 155401  (2002).
\bibitem{Simon2003} S. H. Simon, E. H. Rezayi, and M. V. Milovanovic, Phys. Rev. Lett. \textbf{91}, 046803 (2003).
\bibitem{Wang2003} D.-W. Wang, E. Demler, and S. Das Sarma, Phys. Rev. B \textbf{68}, 165303 (2003).
\bibitem{Doretto2006} R. L. Doretto, A. O. Caldeira, and C. M. Smith, Phys. Rev. Lett. \textbf{97}, 186401 (2006); R. L. Doretto, C. Morais Smith, and A. O. Caldeira, Phys. Rev. B \textbf{86}, 035326 (2012).
\bibitem{Alicea2009}J. Alicea, O. I. Motrunich, G. Refael, and Matthew P. A. Fisher, Phys. Rev. Lett. \textbf{103}, 256403 (2009).
\bibitem{Cipri2014}R. Cipri and N. E. Bonesteel,Phys. Rev. B \textbf{89}, 085109(2014).
\bibitem{Isobe2016}H. Isobe and L. Fu, Phys. Rev. Lett. \textbf{118}, 166401 (2017).
 \bibitem{Sodemann2016}I. Sodemann, I. Kimchi, C. Wang, and T. Senthil, Phys. Rev. B \textbf{95}, 085135 (2017).
\bibitem{Potter2016}A. C. Potter, C. Wang, M. A. Metlitski, and A. Vishwanath, arXiv:1609.08618 (2016).
 \bibitem{Lian2017} B. Lian, S.-C. Zhang, Phys. Rev. Lett. \textbf{120}, 077601 (2018).
 \bibitem{You2017} Yizhi You,	arXiv:1704.03463; Yizhi You, Phys. Rev. B \textbf{97}, 165115(2018).
 \bibitem{Milovanovic2017} M. V. Milovanovi\'c, Phys. Rev. B \textbf{95}, 235304 (2017).


\bibitem{Yoshioka1989}D. Yoshioka, A. H. MacDonald, and S. M. Girvin,Phys. Rev. B \textbf{39}, 1932 (1989).
\bibitem{Spielman2000}  I. B. Spielman, J. P. Eisenstein, L. N. Pfeiffer, and K. W. West, Phys. Rev. Lett. \textbf{84}, 5808 (2000).
\bibitem{Tutuc2004}E. Tutuc, M. Shayegan, and D. A. Huse, Phys. Rev. Lett. \textbf{93}, 036802 (2004).


 \bibitem{Halperin1993} B. I. Halperin, P. A. Lee, and N. Read, Phys. Rev. B \textbf{47}, 7312 (1993).

\bibitem{Read2000} N. Read and D. Green, Phys. Rev. B \textbf{61}, 10267 (2000).

\bibitem{Jain2000}V. W. Scarola, K. Park  and  J. K. Jain, Nature \textbf{406}, 863 (2000)

\bibitem{Brey2000} L. Brey and H. A. Fertig, Phys. Rev. B \textbf{62}, 10268 (2000).

\bibitem{Shi2008}C. Shi, S. Jolad, N. Regnault, and J. K. Jain, Phys. Rev. B \textbf{77}, 155127 (2008).



%
\bibitem{Haldane1985} F. D. M. Haldane, Phys. Rev. Lett. \textbf{55}, 2095 (1985).

\bibitem{Landau1958} L. D. Landau and E. M. Lifshitz, {\it Statistical Physics} (Pergamon, Oxford, 1958).
\bibitem{Senthil2004} T. Senthil, A. Vishwanath, L. Balents, S. Sachdev, and M. P. A. Fisher, Science {\bf303}, 1490 (2004).

\bibitem{Kitaev2006} A. Kitaev, Annals of Physics {\bf321}, 2 (2006).

 \bibitem{MacDonald1990}A. H. MacDonald, P. M. Platzman, and G. S. Boebinger, Phys. Rev. Lett. \textbf{65}, 775 (1990).

 \bibitem{Thouless1982}D. J. Thouless, M. Kohmoto, M. P. Nightingale, and M. den Nijs, Phys. Rev. Lett. \textbf{49}, 405 (1982).
 \bibitem{Niu1985} Q. Niu, D. J. Thouless, and Y.-S. Wu, Phys. Rev. B 31 3372 (1985).
\bibitem{Wen1991}X.G. Wen and A. Zee, Phys. Rev. B \textbf{44}, 274 (1991).
\bibitem{Yang2001}K. Y. and A. H. MacDonald, Phys. Rev. B \textbf{63}, 073301 (2001).
\bibitem{Sheng2006} D. N. Sheng, Z.-Y. Weng, L. Sheng, and F. D. M. Haldane, Phys. Rev. Lett. \textbf{97}, 036808 (2006).
\bibitem{Sheng2011} D. N. Sheng, Z.-C. Gu, K. Sun, and L. Sheng, Nat. Commun. \textbf{2}, 389 (2011).
%


\bibitem{Zanardi2006}P. Zanardi and N. Paunkovi\'{c}, Phys. Rev. E \textbf{74}, 031123 (2006).
\bibitem{Gu2010}S.-J. Gu, International Journal of Modern Physics B \textbf{24}, 4371 (2010).

\bibitem{Bonderson2013} P. Bonderson, C. Nayak, and X.-L. Qi, Journal of Statistical Mechanics: Theory and Experiment 2013.09 (2013): P09016.
\bibitem{Chen2014} X. Chen, L. Fidkowski, and A. Vishwanath, Phys. Rev. B {\bf89},165132 (2014).
\bibitem{Seiberg2016} N. Seiberg and E. Witten, Progress of Theoretical and Experimental Physics 2016.12 (2016).

\bibitem{Wen1995} X.-G. Wen, Advances in Physics {\bf44}, 405 (1995).
\bibitem{Wen2004} X.-G. Wen, {\it Quantum field theory of many-body systems: from the origin of sound to an origin of light and electrons}, Oxford University Press on Demand (2004).

\bibitem{Alicea2012} J. Alicea, Reports on progress in physics {\bf75}, 076501 (2012).

\bibitem{Peskin1978} M. E. Peskin, Annals Phys. {\bf113}, 122 (1978).
\bibitem{Halperin1981} C. Dasgupta and B. I. Halperin, Phys. Rev. Lett. {\bf47}, 1556 (1981).
\end{thebibliography}
\end{document}